\documentclass[pra,twocolumn,showpacs,showkeys]{revtex4}
\usepackage{amsmath}
\usepackage{amssymb}
\usepackage{dsfont}

\newtheorem{theorem}{Theorem} 
\newtheorem{corollary}{Corollary}

\newcommand{\ket}[1]{\left\vert#1\right\rangle}     
\newcommand{\abs}[1]{\left\vert#1\right\vert}   

\begin{document}

\title{Transformations of $W$-Type Entangled States}

\author{S. K{\i}nta\c{s}}   
\author{S. Turgut}
\affiliation{ Department of Physics, Middle East Technical University,\\
TR-06531, Ankara, Turkey}

\begin{abstract}
The transformations of $W$-type entangled states by using local
operations assisted with classical communication are investigated.
For this purpose, a parametrization of the $W$-type states which
remains invariant under local unitary transformations is proposed
and a complete characterization of the local operations carried
out by a single party is given. These are used for deriving the
necessary and sufficient conditions for deterministic
transformations. A convenient upper bound for the maximum
probability of distillation of arbitrary target states is also
found.
\end{abstract}



\maketitle

\section{Introduction}

Entanglement can be regarded as a nonlocal resource which enables
us to achieve some classically impossible tasks such as dense
coding\cite{aBennett} and teleportation\cite{bBennett}. However,
such tasks can be accomplished only when certain special entangled
states have already been established between particles distributed
to remote parties. As the transfer of such particles over noisy
quantum channels results in the decoherence of the state, the
transformation of entangled states using only local operations
assisted with classical communication (LOCC) has arisen as an
important problem in quantum information theory. Such
transformations also form an operational basis for
quantifying\cite{cBennett,VidalMonotone} and distinguishing the
different types\cite{SLOCC1,SLOCC2} of entanglement.
Alternatively, the deterministic transformations can be used for
defining a natural partial order between the states\cite{Nielsen},
which provides another approach for assessing the entanglement
content of the states.

Nearly all aspects of the transformations of pure bipartite states
have been understood\cite{cBennett,Nielsen,Lo,Vidal,Jonathan}. The
existence of the Schmidt decomposition of such states appears to
be immensely useful in the analysis of the associated
transformations. A particularly convenient aspect of the
decomposition is the existence of a one-to-one correspondence
between the equivalence classes of states under local unitary (LU)
transformations and the unique Schmidt coefficients of the state.
As a result of this, all statements about the transformations of
bipartite pure states can be expressed entirely in terms of the
Schmidt coefficients.

The absence of a simple and powerful representation like the
Schmidt decomposition has been a great impediment in the analysis
of the transformations of multipartite entangled states. For this
reason, studies on such transformations are focused on specific
types of states. For transformations between multipartite states
that have a generalized Schmidt decomposition, it is found that
the results obtained for bipartite states can be directly applied
without changes\cite{Xin_Duan}. The two special types of genuine
multipartite states of three qubits generally attracts a special
interest, and for this reason various aspects of the
transformations of Greenberger-Horne-Zeilinger (GHZ) type
states\cite{Acin,Spedalieri,Lo_GHZ,Turgut} and $W$-type
states\cite{Cao1,Cao2,Fortescue1,Fortescue2} are investigated.
Most of these studies are concerned with special source or target
states, however, and lack a systematic analysis of the
transformations. Such a systematic investigation not only enables
us to better assess the order between the states in terms of their
entanglement content, but also provides a framework upon which
future studies can be built. A systematic treatment of the
transformations of GHZ-type states has recently been given
\cite{Turgut} and the purpose of the current article is to do the
same for the transformations of the $W$-type states.

The organization of the article is as follows. In Sec.
\ref{sec:param}, the $W$-type states are defined and a convenient
parametrization of these states is given. The LU equivalence
relation between states is also described in this section. In Sec.
\ref{sec:locop}, a complete characterization of the local quantum
operations carried out by a single party is given.
Sec.~\ref{sec:det} contains two particular applications: complete
characterization of the deterministic transformations and finding
an upper bound for the maximum distillation probability for
arbitrary states.

\section{States and their Parametrization\label{sec:param}}

Standard $W$ state of $p$ qubits ($p\geq3$) distributed to $p$ different
parties is given by
\begin{equation}
\ket{W} = \frac{1}{\sqrt{p}} \left(
  \ket{10\cdots0}+\ket{01\cdots0}+ \cdots + \ket{00\cdots1}
  \right)~,
\end{equation}
where the labels in kets denote states of the particles
$1,2,\ldots,p$, in that order. A state $\ket{\Psi}$ of $p$
particles will be called a \emph{$W$-type state}, if it is
stochastically reducible\cite{LUpaper} from the state $\ket{W}$ by
LOCC. Note that, by the term $W$-type, we also embrace states
where some parties are unentangled from the rest. Especially, the
product states and bipartite entangled states with Schmidt rank 2
are covered by this definition. Since this set of states is closed
under stochastic reducibility relation by definition, any state in
this set can only be transformed to states in the same set. Hence,
all possible local manipulations of these states are within the
scope of the current work.

If $\ket{\Psi}$ is a $W$-type state, then there are local
operators $A_k$ such that $\ket{\Psi}=(A_1\otimes\cdots\otimes
A_p)\ket{W}$ and therefore it can be expressed as
\begin{equation}
  \ket{\Psi} = \sum_{k=1}^p \ket{u_1\otimes \cdots \otimes
  u_{k-1}\otimes v_k\otimes u_{k+1}\otimes\cdots u_p}
\end{equation}
where $\ket{u_k}$ and $\ket{v_k}$ are vectors (which need to be
neither normalized nor orthogonal) in the Hilbert space
$\mathcal{H}_k$ of the $k$th particle. Since only two vectors are
involved for each party, it can always be assumed that each
particle is qubit and $\dim\mathcal{H}_k=2$.

For the purpose of obtaining a unique representation, orthonormal
sets $\{\ket{\alpha_{k}},\ket{\beta_k}\}$ can be defined in
$\mathcal{H}_k$ in such a way that $\ket{\alpha_k}$ is parallel to
$\ket{u_k}$ and hence
\begin{eqnarray}
  \ket{u_k} &=& c_k \ket{\alpha_k}  \quad, \\
  \ket{v_k} &=& c^\prime_k \ket{\alpha_k}+c^{\prime\prime}_k \ket{\beta_k}\quad,
\end{eqnarray}
for some expansion constants. Expanding $\ket{\Psi}$ in these
bases we get
\begin{eqnarray}
\ket{\Psi} = z_0\ket{\alpha_1 \cdots \alpha_{p}}
+ \sum_{k=1}^p z_k
          \ket{\alpha_1 \cdots \alpha_{k-1} \beta_k \alpha_{k+1} \cdots \alpha_p}
\end{eqnarray}
for some complex coefficients $z_0,z_1,\ldots,z_p$. Finally, the
phases of these basis vectors can be redefined so that all
expansion coefficients are nonnegative real numbers. With this
redefinition, the state becomes
\begin{eqnarray}
\ket{\Psi} &=& \sqrt{x_{0}}\ket{\alpha_1 \cdots \alpha_{p}}
                \label{eq:xrepr} \\
         & & + \sum_{k=1}^p \sqrt{x_k} \ket{\alpha_1\cdots \alpha_{k-1} \beta_k \alpha_{k+1} \cdots \alpha_p}~~,
               \nonumber
\end{eqnarray}
where $x_k=\abs{z_k}^2$. In short, it is shown that for any
$W$-type state there are nonnegative real numbers,
$x_0,x_1,\cdots,x_p$ and local orthonormal bases
$\{\ket{\alpha_{k}},\ket{\beta_k}\}$ in $\mathcal{H}_k$ such that
the expansion above is valid.

Note that we have
\begin{equation}
    x_0 + \sum_{k = 1}^p   x_k = 1
    \label{eq:x0_function}
\end{equation}
by normalization. The vector $\mathbf{x}=(x_1,x_2,\ldots,x_p)$ is
the basic mathematical object that will be used in describing the
$W$-type states. The zeroth component $x_0$ is not considered as
an independent component; it will rather be considered as a
function of the vector $\mathbf{x}$ given by Eq.~
(\ref{eq:x0_function}). For distinguishing from the zeroth, all
other numbers $x_1,\ldots,x_p$ will be called \emph{party
components}. All possible allowed vectors $\mathbf{x}$ form a
subset $\mathcal{S}$ of $\mathbb{R}^p$ which is actually a simplex
defined by $x_k\geq0$ for $k=1,2,\ldots,p$ and
$x_1+\cdots+x_p\leq1$.

An LU transformation of the state $\ket{\Psi}$ changes only the
local orthonormal bases $\{\ket{\alpha_{k}},\ket{\beta_k}\}$.
Therefore the vector parameter $\mathbf{x}$ is invariant under LU
transformations, which makes it a good candidate for parametrizing
LU equivalence classes. The state $\ket{\Psi}$ given in
Eq.~(\ref{eq:xrepr}) is LU equivalent to the following
representative state
\begin{eqnarray}
  \ket{\Phi(\mathbf{x})}  &=& \sqrt{x_0}\ket{000\cdots0}
                                 +\sqrt{x_1}\ket{100\cdots0}   \\
          & & +  \sqrt{x_2}\ket{010\cdots0}+\cdots+\sqrt{x_p}\ket{000\cdots1} \nonumber\\
    &=& \sqrt{x_0}\ket{\mathbf{0}}+\sum_{k=1}^p\sqrt{x_k}\ket{\mathbf{1}_k}
\end{eqnarray}
where $\ket{\mathbf{0}}$ is shorthand for the state where all
qubits are in $0$ state and $\ket{\mathbf{1}_k}$ represents the
state where the $k$th qubit is in $1$ state and all the others are
in $0$ state.

For some states $\ket{\Psi}$, there might be different
representations of the form (\ref{eq:xrepr}), i.e., the state
might be associated with two different parameter vectors
$\mathbf{x}$ and $\mathbf{x}^\prime$. This is equivalent to saying
that the vectors $\ket{\Phi(\mathbf{x})}$ and
$\ket{\Phi(\mathbf{x}^\prime)}$ are LU equivalent. In such a case,
we will say that the vectors $\mathbf{x}$ and $\mathbf{x}^\prime$
are equivalent and show this relation by
$\mathbf{x}\sim\mathbf{x}^\prime$. With the complete
characterization of this equivalence relation, the simplex
$\mathcal{S}$ becomes a natural working ground in the
investigation of the transformation of $W$-type states.
Consequently, we will usually talk about the transformations of
points in $\mathcal{S}$. For example, we say that $\mathbf{x}$ can
be transformed deterministically to $\mathbf{y}$, when it is
actually meant that a state associated with the vector
$\mathbf{x}$ can be deterministically transformed to another state
that has parameter vector $\mathbf{y}$.

For investigating the equivalence relation in the parameter space
$\mathcal{S}$, it is useful to utilize quantities that do not
depend on the particular representation used for the state. One
such property is the concurrence\cite{Wootters_Concur}
corresponding to the bipartite entanglement between a subset of
the parties and the rest. Let $\mathcal{C}_k$ be the concurrence
of the entanglement of the $k$th party with the rest. It is given
by
\begin{equation}
 \mathcal{C}_k = 2\sqrt{\det\rho^{(k)}} =2\sqrt{x_k(1-x_0-x_k)}
\end{equation}
where $\rho^{(k)}$ is the reduced density matrix of the $k$th
particle. The corresponding quantity for a subset
$R=\{k_1,k_2,\ldots,k_r\}$ of parties can be computed easily with
a simple trick. Note that the state $\ket{\Psi}$ will still be
classified as a $W$-type state when all parties in $R$ is
reinterpreted as a single party. In that case, the expansion in
Eq.~(\ref{eq:xrepr}) is altered by taking the $x$-parameter
corresponding to $R$ as
\begin{equation}
  x_R=x_{k_1}+x_{k_2}+\cdots+x_{k_r}~~,
  \label{eq:x_aggregate}
\end{equation}
and by taking the associated local orthonormal basis as
$\{\ket{A_R},\ket{B_R}\}$ where
\begin{eqnarray}
\ket{A_R} &=& \ket{\alpha_{k_1}\otimes \alpha_{k_2}\otimes\cdots\otimes \alpha_{k_r}}~, \\
\ket{B_R} &=& \sum_{j=1}^r  \sqrt{\frac{x_{k_j}}{x_R}}
              \ket{\alpha_{k_1}\otimes\cdots\otimes\beta_{k_j}\otimes\cdots\otimes \alpha_{k_r}}.
\end{eqnarray}
As a result, the concurrence of the entanglement between $R$ and
the rest of the parties is given by
$\mathcal{C}_R=2\sqrt{x_R(1-x_0-x_R)}$.

For any pair of distinct parties $k$ and $\ell$, define the
following quantity
\begin{equation}
\mathcal{D}_{k\ell} = \frac{1}{8}(\mathcal{C}_k^2 + \mathcal{C}_\ell^2 - \mathcal{C}_{k\ell}^2)~~.
\end{equation}
Note that $\mathcal{D}_{k\ell}$ does not depend on the way
$\ket{\Psi}$ is represented. Moreover, it is related to the
parameter vector $\mathbf{x}$ by $\mathcal{D}_{k\ell}=x_k x_\ell$.
This shows that if $\mathbf{x}\sim\mathbf{x}^\prime$ then
$x_kx_\ell=x_k^\prime x_\ell^\prime$ for all distinct pairs of
parties $k$ and $\ell$.

The last relation implies that when $\ket{\Psi}$ has a
representation (\ref{eq:xrepr}) for which at least two components
of $\mathbf{x}$ is nonzero, then all other party-components are
unique. In particular, if $x_r$ and $x_s$ are nonzero, then for
any party $k\neq r,s$, the coefficient $x_k$ is given as
\begin{equation}
  x_k = \sqrt{\frac{\mathcal{D}_{kr}\mathcal{D}_{ks}}{\mathcal{D}_{rs}}}~~
\end{equation}
and hence such $x_k$ are unique. Consequently, when three
components of $\mathbf{x}$ are nonzero, then every component of
$\mathbf{x}$ is unique.

The equivalence relation in $\mathcal{S}$ and the type of
entanglement that these vectors represent can be completely
described as follows. Let $\ket{\Psi}$ be a $W$-type state having
a parameter vector $\mathbf{x}$. There are three possibilities
depending on the number of nonzero party components of
$\mathbf{x}$.
\begin{itemize}
\item[(i)] At least three party-components of $\mathbf{x}$ are nonzero:
    in this case, the representation in Eq.~(\ref{eq:xrepr}) is unique;
    in other words $\mathbf{x}\sim\mathbf{x}^\prime$ if and only if $\mathbf{x}=\mathbf{x}^\prime$.
    Moreover, $\ket{\Psi}$ is a \emph{truly multipartite} state
    (i.e., at least three parties are entangled with each other) and a
    given party $k$ is entangled with the rest ($\mathcal{C}_k\neq0$)
    if and only if $x_k\neq0$.
\item[(ii)] Only two party-components of $\mathbf{x}$ are nonzero, say
    $x_r,x_s\neq0$: in that case the vector parameter is not
    unique in general\footnote{The vector parameter $\mathbf{x}$ for such bipartite
    states is unique if and only if $x_r=x_s=1/2$, i.e., the state is
    a maximally entangled bipartite state.}.
    The state $\ket{\Psi}$ is a bipartite
    entangled state between $r$th and $s$th particles having the
    concurrence $\mathcal{C}=2\sqrt{x_rx_s}$. We have
    $\mathbf{x}\sim\mathbf{x}^\prime$ if and only if $x_rx_s=x_r^\prime x_s^\prime$
    and $x_k^\prime=0$ for all $k\neq0,r,s$.
\item[(iii)] Finally, in all the other cases, i.e., when all
    components of $\mathbf{x}$ are zero or if only one of them is nonzero,
    then $\ket{\Psi}$ is a product state. Obviously,
    $\mathbf{x}\sim\mathbf{x}^\prime$ if and only if
    $\mathbf{x}^\prime$ satisfies the same property.
\end{itemize}

When a parameter vector $\mathbf{x}$ is unique, then the basis
vectors used in the representation in Eq.~(\ref{eq:xrepr}) are
also unique. This can be seen as follows. First, note that if
$x_k=0$ for some $k$, then the $k$th particle is unentangled from
the rest and therefore only $\ket{\alpha_k}$ is uniquely defined
(up to a phase) and the vector $\ket{\beta_k}$ is irrelevant. For
two distinct parties $k$ and $\ell$ for which
$\mathcal{C}_{k\ell}\neq0$ (i.e., these two parties are entangled
with the rest), consider the reduced density matrix
$\rho^{(k\ell)}$ on the Hilbert space
$\mathcal{H}_k\otimes\mathcal{H}_\ell$ of the corresponding
particles. This is an operator on a $4$-dimensional space having
rank $2$. Hence, its eigensubspace corresponding to zero
eigenvalue is $2$-dimensional and spanned by the following two
vectors
\begin{eqnarray}
 \ket{\Theta_1^{(k,\ell)}} &=& \ket{\beta_k\otimes\beta_\ell} \quad,\\
 \ket{\Theta_2^{(k,\ell)}} &=& \sqrt{x_\ell}\ket{\beta_k\otimes\alpha_\ell}- \sqrt{x_k}\ket{\alpha_k\otimes\beta_\ell}\quad.
\end{eqnarray}
It is clear that, if both $x_k$ and $x_\ell$ are nonzero, then
this eigensubspace contains only one vector (direction) in product
form, namely $\ket{\Theta_1^{(k,\ell)}}$. This enables us to
define $\ket{\beta_k}$ and $\ket{\beta_\ell}$ uniquely up to a
phase. By orthogonality, the vectors $\ket{\alpha_k}$ and
$\ket{\alpha_\ell}$ are also unique. In summary, when the $W$-type
state is truly multipartite, then the basis vectors are also
uniquely defined.

\section{Local Operations by One Party\label{sec:locop}}

In order to be able to analyze the transformations of $W$-type states,
one should first describe the local quantum operations carried out
by a single party and its effect on the parameters. This section
is concerned with this description and its immediate implications.
The main result is given by the following theorem.

\begin{theorem}
\label{thm:locop}
Suppose that the $k$th party carries out a local operation on a
$W$-type state with parameter vector $\mathbf{x}$. The set of
final states with vectors $\mathbf{x}_\lambda$ can be produced
with probabilities $P_\lambda$ if and only if for each possible
outcome $\lambda$ there are vectors
$\mathbf{x}^\prime_\lambda\sim\mathbf{x}_\lambda$ and there are
nonnegative scale factors $s_\lambda$ such that
\begin{itemize}
\item[(i)] $x^\prime_{\lambda,\ell}=s_\lambda x_\ell$ for all
    $\ell\neq k,0$;
\item[(ii)] $\sum_\lambda P_\lambda s_\lambda = 1$ and
\item[(iii)]
\begin{equation}
      \sum_\lambda P_\lambda \sqrt{s_\lambda x^\prime_{\lambda,0}} \geq \sqrt{x_0} ~~.
      \label{eq:polygonal_ineq}
\end{equation}
\end{itemize}
\end{theorem}

Note that the requirement of finding a new vector
$\mathbf{x}^\prime_\lambda$ may become necessary only if the final
state $\mathbf{x}_\lambda$ corresponds to a bipartite or a product
state. Otherwise, if $\mathbf{x}_\lambda$ corresponds to a truly
multipartite final state, we necessarily have
$\mathbf{x}^\prime_\lambda=\mathbf{x}_\lambda$. Note also that, it
is not necessary that the initial state $\mathbf{x}$ is truly
multipartite; the theorem is valid for all the other cases as
well.

We first start proving the necessity of the conditions listed.
Suppose that the $k$th party carries out a local operation on the
state $\ket{\Phi(\mathbf{x})}$ and let $\{M_\lambda\}$ be the set
of measurement operators that describe this operation. They
satisfy the normalization relation $\sum_\lambda M_\lambda^\dagger
M_\lambda=\mathds{1}_k$ where $\mathds{1}_k$ denotes the identity
operator on $\mathcal{H}_k$. For each outcome $\lambda$, an
orthonormal basis $\{\ket{a_\lambda},\ket{b_\lambda}\}$ can be
found in $\mathcal{H}_k$ such that
\begin{eqnarray}
  M_\lambda \ket{0} &=& A_\lambda \ket{a_\lambda} \quad,\\
  M_\lambda \ket{1} &=& B_\lambda \ket{a_\lambda}+C_\lambda\ket{b_\lambda} \quad,\\
\end{eqnarray}
where $A_\lambda$ and $C_\lambda$ are nonnegative real numbers and
$B_\lambda$ is some complex number. Note that the basis
$\{\ket{a_\lambda},\ket{b_\lambda}\}$ depends only on a local
unitary transformation that may be applied by the $k$th party
after the outcome $\lambda$ is obtained and hence it is
irrelevant. This basis will be chosen to be $\{\ket{0},\ket{1}\}$
for simplifying the notation below. The normalization relation of
the measurement operators can be expressed as
\begin{equation}
  \sum_\lambda A_\lambda^2 = \sum_\lambda  \abs{B_\lambda}^2+C_\lambda^2 =1
  \quad,\quad
  \sum_\lambda A_\lambda B_\lambda = 0\quad.
  \label{eq:ABC_normalization}
\end{equation}

When the outcome $\lambda$ is obtained, the final 
state is
\begin{eqnarray}
  (M_\lambda \otimes\mathds{1}_k^\prime) \ket{\Phi(\mathbf{x})}
     &=& (A_\lambda \sqrt{x_0}+B_\lambda \sqrt{x_k})\ket{\mathbf{0}}     \\
     & & + \sum_{\ell\neq k}A_\lambda \sqrt{x_\ell}\ket{\mathbf{1}_\ell}+C_\lambda\sqrt{x_k}\ket{\mathbf{1}_k}  \nonumber
\end{eqnarray}
where $\mathds{1}_k^\prime$ represents the identity operator on
all particles except the $k$th. The probability of the outcome is
given by the square of the norm of the vector above
\begin{equation}
  P_\lambda = \abs{A_\lambda \sqrt{x_0}+B_\lambda\sqrt{x_k}}^2 + A_\lambda^2(1-x_0-x_k)+ C_\lambda^2 x_k~~,
\end{equation}
and after an appropriate phase redefinition of the local basis
vectors of the $k$th qubit, the vector parameter
$\mathbf{x}_\lambda$ of the final state can be found as
\begin{eqnarray}
  x_{\lambda,k} &=& \frac{C_\lambda^2 x_k}{P_\lambda} \quad,  \\
  x_{\lambda,\ell}  &=& \frac{A_\lambda^2 x_\ell}{P_\lambda} ~~(\ell\neq0,k) \quad,\\
  x_{\lambda,0} &=& \frac{\abs{A_\lambda \sqrt{x_0}+B_\lambda\sqrt{x_k}}^2}{P_\lambda}  \quad.
\end{eqnarray}
Obviously, $s_\lambda=A_\lambda^2/P_\lambda$ and it can be
immediately seen that the conditions (i) and (ii) in the theorem
is satisfied. For proving condition (iii), note that
\begin{eqnarray}
\sum_\lambda P_\lambda \sqrt{s_\lambda x_{\lambda,0}}
   &=& \sum_\lambda A_\lambda \abs{A_\lambda \sqrt{x_0}+B_\lambda\sqrt{x_k}}  \nonumber \\
   &\geq& \sum_\lambda A_\lambda \left(A_\lambda \sqrt{x_0}+\mathrm{Re}(B_\lambda)\sqrt{x_k}\right)\nonumber \\
   &=& \sqrt{x_0}~~,
\end{eqnarray}
where we have used the fact that the modulus of a complex number
is greater than its real part in order to introduce the inequality
and then invoked Eq.~(\ref{eq:ABC_normalization}). This completes
the proof of the necessity of the conditions.

In order to prove the sufficiency, first suppose that the
conditions (i), (ii) and (iii) are satisfied. Let $\lambda_0$ be
the outcome which contributes the largest term in the summation in
Eq.~(\ref{eq:polygonal_ineq}). If it happens that there is only
one possible outcome (i.e., $P_{\lambda_0}=1$) or if
\begin{equation}
 P_{\lambda_0} \sqrt{s_{\lambda_0} x^\prime_{\lambda_0,0}}
   >
 \sum_{\lambda\neq\lambda_0} P_\lambda \sqrt{s_\lambda x^\prime_{\lambda,0}}
\end{equation}
then redefine the outcomes so that $\mathbf{x}_{\lambda_0}$
appears twice in the result set with probabilities
$P_{\lambda_0}/2$ each. In that case,
Eq.~(\ref{eq:polygonal_ineq}) can be interpreted as the polygonal
inequality, i.e., the generalization of the triangle inequality to
polygons. Imagining the corresponding polygon on the complex
plane, we can find angles $\phi_\lambda$ such that the equality
\begin{equation}
  \sum_\lambda P_\lambda \sqrt{s_\lambda x^\prime_{\lambda,0}} e^{i\phi_\lambda}= \sqrt{x_0}
\end{equation}
is satisfied. In that case, it is possible to go backwards in the
derivation given above by defining
\begin{eqnarray}
  A_\lambda   &=&   \sqrt{P_\lambda s_\lambda}   \quad,  \\
  B_\lambda   &=& \sqrt{\frac{P_\lambda}{x_k}}\left(\sqrt{x_{\lambda,0}^\prime}e^{i\phi_\lambda}-\sqrt{s_\lambda x_0}\right) \quad,  \\
  C_\lambda   &=& \sqrt{\frac{P_\lambda x_{\lambda,k}^\prime}{x_k}}   \quad,   \\
  M_\lambda   &=& \left[\begin{array}{cc} A_\lambda & B_\lambda \\ 0 & C_\lambda \end{array}\right]\quad.
\end{eqnarray}
It is straightforward to see that the relations
(\ref{eq:ABC_normalization}) are satisfied and hence
$\{M_\lambda\}$ satisfy the normalization condition of measurement
operators. It is also straightforward to check that the final
state $\mathbf{x}^\prime_\lambda$ is produced with probability
$\lambda$. This completes the proof of the theorem.$\Box$

Let us investigate some implications of the conditions in the
theorem. First, applying the Schwarz inequality to condition (iii)
of the theorem gives
\begin{equation}
  \sqrt{x_0}\leq \sum_\lambda P_\lambda \sqrt{s_\lambda x^\prime_{\lambda,0}}
    \leq \sqrt{\sum_\lambda P_\lambda s_\lambda}\sqrt{\sum_\lambda P_\lambda x^\prime_{\lambda,0}}
\end{equation}
which shows that
\begin{equation}
  x_0 \leq \sum_\lambda P_\lambda x^\prime_{\lambda,0}~~,
  \label{ineq:comp_0}
\end{equation}
with equality holding if and only if $x^\prime_{\lambda,0}$ is
proportional to $s_\lambda$. In short, the zeroth component of the
parameter vector is nondecreasing on the average. Together with
the condition (i), the following inequalities can be written for
the party components of the vectors.
\begin{eqnarray}
  x_\ell &=& \sum_\lambda P_\lambda x^\prime_{\lambda,\ell}   \quad(\ell\neq0,k),
        \label{ineq:comp_ell} \\
  x_k &\geq& \sum_\lambda P_\lambda x^\prime_{\lambda,k}~~,
        \label{ineq:comp_k}
\end{eqnarray}
i.e., the $k$th component of the parameter vector is
non-increasing on the average, while the averages of all the other
components do not change. These inequalities imply that, for any
to distinct parties $k$ and $\ell$, the function
$\sqrt{x_kx_\ell}$, which depends only on the LU equivalence class
of the state, is an entanglement monotone\cite{VidalMonotone}.

An important corollary that can be deduced from the theorem is concerned
with the deterministic transformation induced only by the local operations
of the $k$th party.
\begin{corollary}
\label{cor:loc_det}
Let $\mathbf{x}$ and $\mathbf{y}$ be two vectors in $\mathcal{S}$
that differ only in their $k$th component such that $x_k>y_k$.
Then, the state $\ket{\Phi(\mathbf{x})}$ can be transformed to
$\ket{\Phi(\mathbf{y})}$ by the local operations of the $k$th
party only.
\end{corollary}
The corollary follows straightforwardly from Theorem
\ref{thm:locop} as there is only one outcome $\lambda$;
necessarily $s_\lambda=1$ and finally $y_0>x_0$. Hence, all three
conditions of the theorem are satisfied. A two-outcome general
measurement that does the transformation can be constructed easily
by using the reasoning given in the proof of the Theorem
\ref{thm:locop}. Note that, in order to bring the state to the
final form $\ket{\Phi(\mathbf{y})}$, additional LU transformations
by all the other parties are also needed.


\section{General LOCC Transformations \label{sec:det}}

The relations in Eq.~(\ref{ineq:comp_0}-\ref{ineq:comp_k}) can be
easily extended to a general protocol where a sequence of local
operations is carried out by different parties. Let the variable
$\Lambda$ be used for labeling the sequence of outcomes obtained
in these operations and let $P_\Lambda$ be the probability for
this outcome. Let $\mathbf{x}$ be the parameter vector
corresponding to the initial state and let
$\mathbf{x}_\Lambda^\prime$ denote the vector, whose existence is
guaranteed by Theorem \ref{thm:locop}, that corresponds to the
final state when the outcome $\Lambda$ is obtained. Using
Eq.~(\ref{ineq:comp_ell},\ref{ineq:comp_k}) the following
inequality can be seen to be valid
\begin{equation}
  x_\ell \geq \sum_\Lambda P_\Lambda x_{\Lambda,\ell}^\prime \quad(\ell=1,2,\ldots,p)~.
  \label{eq:av_state_ineq1}
\end{equation}
Obviously, if the $\ell$th party never does a local operation,
then equality sign holds in Eq.~(\ref{eq:av_state_ineq1}).

At this point, it is very convenient to introduce a partial order
between the vectors in $\mathcal{S}$. For two such vectors
$\mathbf{x}$ and $\mathbf{y}$, we will say that $\mathbf{x} \geq
\mathbf{y}$ if and only if $x_\ell \geq y_\ell$ for all parties
$\ell=1,2,\ldots,p$. Using this relation,
Eq.~(\ref{eq:av_state_ineq1}) can be expressed more compactly as
\begin{equation}
  \mathbf{x} \geq \sum_\Lambda P_\Lambda \mathbf{x}^\prime_\Lambda ~~.
  \label{eq:av_state_ineq2}
\end{equation}
Therefore, during all the intermediate moments of the whole
transformation process, the average of the state vectors follows a
monotonically decreasing path in $\mathcal{S}$. It turns out that
this partial order completely describes the deterministic
transformations.

\begin{theorem}
\label{thm:det}
A $W$-type state with parameter vector $\mathbf{x}$ can be transformed by
LOCC to another state with vector $\mathbf{y}$ if and only if there is
$\mathbf{y}^\prime\sim\mathbf{y}$ such that
$\mathbf{x}\geq\mathbf{y}^\prime$.
\end{theorem}

Again, the need for using another final state vector
$\mathbf{y}^\prime$ may become necessary only when $\mathbf{y}$
corresponds to a bipartite state. The proof of the theorem is
rather straightforward. The sufficiency part is already partially
covered by the Corollary \ref{cor:loc_det}. Hence, if
$\mathbf{x}\geq\mathbf{y}^\prime$, then every party carries out a
local operation whose sole effect is the decrease of the
corresponding component of the parameter vector. In other words,
for all $k$, $k$th party changes the $k$th component of the vector
from $x_k$ to $y_k^\prime$.

For the sufficiency part of the proof, suppose that $\mathbf{x}$
to $\mathbf{y}$ conversion is possible. If $\mathbf{y}$ is a truly
multipartite state, then Eq.~(\ref{eq:av_state_ineq2}) already
gives us $\mathbf{x}\geq\mathbf{y}$ and there is nothing further
to do. On the other hand, if $\mathbf{y}$ corresponds to a
bipartite state, then different final state vectors
$\mathbf{x}^\prime_\Lambda$ may appear in
Eq.~(\ref{eq:av_state_ineq2}) in which case the following analysis
has to be made. Let the final state be a bipartite entangled state
between the parties $r$ and $s$. Let $\bar{x}_r^\prime$ and
$\bar{x}_s^\prime$ represent the average of the $r$th and $s$th
components of the final vectors $\mathbf{x}_\Lambda^\prime$. By
Eq.~(\ref{eq:av_state_ineq2}) we have
\begin{eqnarray}
  x_r &\geq& \bar{x}_r^\prime = \sum_\Lambda P_\Lambda x^\prime_{\Lambda,r}\quad,\\
  x_s &\geq& \bar{x}_s^\prime = \sum_\Lambda P_\Lambda x^\prime_{\Lambda,s}\quad.
\end{eqnarray}
On the other hand $\mathbf{x}^\prime_\Lambda\sim\mathbf{y}$ and hence we
have $x^\prime_{\Lambda,r}x^\prime_{\Lambda,s}=\mathcal{C}^2/4$ where
$\mathcal{C}=2\sqrt{y_ry_s}$ is the concurrence of the final state. We
then use the Schwarz inequality as follows
\begin{eqnarray}
  \frac{\mathcal{C}}{2}  &=&  \sum_\Lambda P_\Lambda \sqrt{x^\prime_{\Lambda,r}x^\prime_{\Lambda,s}} \\
                    &\leq& \sqrt{ \sum_\Lambda P_\Lambda x^\prime_{\Lambda,r} }
                           \sqrt{ \sum_\Lambda P_\Lambda x^\prime_{\Lambda,s} }  \\
                    &=& \sqrt{\bar{x}_r^\prime \bar{x}_s^\prime}~.
\end{eqnarray}
We can now define a suitable $\mathbf{y}^\prime$ vector by
\begin{equation}
  y^\prime_r  =  \bar{x}_r^\prime \quad,\quad
  y^\prime_s  =  \frac{\mathcal{C}^2}{4 \bar{x}_r^\prime} \quad,\quad
  y^\prime_\ell  =  0 \quad (\ell\neq0,r,s)~~.
\end{equation}
It can now be easily verified that
$\mathbf{y}^\prime\sim\mathbf{y}$, $y_r^\prime\leq x_r$ and
finally $y_s^\prime\leq \bar{x}_s^\prime\leq x_s$. Therefore we
have $\mathbf{y}^\prime \leq \mathbf{x}$. This completes the proof
of the theorem. $\Box$

As a result, the order $\geq$ defined on the points of the simplex
$\mathcal{S}$ is closely related to the partial order defined by
the deterministic LOCC convertibility relation. Obviously, there
are pairs of states which cannot be converted into each other,
hence the order is partial. A few words can also be said about the
maximal states. Note that any state having a parameter vector
$\mathbf{x}$ with nonzero zeroth component ($x_0\neq0$) is not
maximally entangled; such a state can always be deterministically
reducible from a different state. The truly multipartite states
that have a vanishing zeroth component ($x_0=0$) are always
maximally entangled. The standard $W$ state with
\begin{equation}
  \mathbf{x}_W = \frac{1}{p} (1,1,\ldots,1)\quad,
\end{equation}
is in this set but there are many more states that are also
maximally entangled.

Next, we turn our attention to the probabilistic transformations.
It appears that the inequality (\ref{eq:av_state_ineq2}) is not
sufficient for a complete description of possible probabilistic
transformations. Still, it can be used for finding suitable bounds
on some problems of interest. For example, it can be utilized for
finding a good upper bound for the maximum probability of
distilling a truly multipartite target state $\mathbf{y}$ from an
initial state $\mathbf{x}$. To do this, consider a transformation
where $\mathbf{x}$ is converted into various states
$\mathbf{x}^\prime_\Lambda$, where the parametrization is such
that Eq.~(\ref{eq:av_state_ineq2}) holds. Let $P$ be the total
probability of outcomes $\Lambda$ where
$\mathbf{x}_\Lambda^\prime=\mathbf{y}$. By
Eq.~(\ref{eq:av_state_ineq2}), we have
\begin{equation}
  P\mathbf{y} + \sum_{ \Lambda (\mathbf{x}^\prime_\Lambda\nsim\mathbf{y}) }    P_\Lambda\mathbf{x}^\prime_\Lambda  \leq  \mathbf{x}~.
  \label{eq:Pmax_bound_derivation}
\end{equation}
Since the sum term gives a vector with nonnegative components, we
have $P\mathbf{y}\leq\mathbf{x}$. Therefore, the probability $P$
is bounded from above by
\begin{equation}
 P \leq \mathcal{P}_{\mathrm{bnd}}(\mathbf{x}\rightarrow\mathbf{y})=\min\left(\frac{x_1}{y_1},\frac{x_2}{y_2},\ldots,\frac{x_p}{y_p},1\right)~~.
 \label{eq:P_bnd}
\end{equation}
Although this bound is tight for some transitions, e.g.,
deterministic transformations, there are some transitions where
the maximum probability and the bound above differ. For example,
consider the case where $\mathbf{x}=t\mathbf{y}$ and $0<t<1$ for
which the bound is
$\mathcal{P}_{\mathrm{bnd}}(t\mathbf{y}\rightarrow\mathbf{y})=t$.
If it were possible that $P=t$, then it would be necessary that
all of the other outcomes are equal to the zero vector,
$\mathbf{x}_\Lambda^\prime=\mathbf{0}$, and
Eq.~(\ref{eq:Pmax_bound_derivation}) is an equality. Therefore,
Eq.~(\ref{ineq:comp_0}) must be an equality as well for all
intermediate local operations. Since the Schwarz inequality has
been used for obtaining Eq.~(\ref{ineq:comp_0}), the equality
conditions of that inequality implies that
$x_{\lambda,0}^\prime\propto s_\lambda$. However, this is in
contradiction with the existence of outcomes
$\mathbf{x}^\prime_\lambda=\mathbf{0}$ for which we have
$s_\lambda=0$, but $x_{\lambda,0}^\prime=1$. This shows that the
distillation probability $P=t$ cannot be reached.

It appears that for cases where the zeroth component of both the
initial and final vectors are zero ($x_0=y_0=0$), the maximum
probability of distillation reaches to the bound in
Eq.~(\ref{eq:P_bnd}). The special case where the final state is
the standard $W$ state (i.e., $\mathbf{y}=\mathbf{x}_W$) has
already been investigated before\cite{Cao1,Cao2} for which a
protocol is proposed which achieves the distillation probability
of $P=p\min(x_1,\ldots,x_p)$. Since this is also equal to
$\mathcal{P}_{\mathrm{bnd}}(\mathbf{x}\rightarrow\mathbf{y})$,
their protocol is optimal.

It appears that their protocol can be directly adapted to all
other transformations on this particular face of the simplex
$\mathcal{S}$, i.e., the face formed by the vectors that have a
vanishing zeroth component. First, note that when all produced
states are on that face, the relation (iii) of Theorem
\ref{thm:locop} is automatically satisfied. This simplifies the
analysis of such transformations considerably. The protocol can be
constructed as follows: Suppose that the ratio $r_k=x_k/y_k$
becomes the minimum for the first $m$ parties, i.e., let
$r_1=\cdots=r_m<r_{m+1},\ldots,r_p$. The local operations are done
by the last $p-m$ parties, in any order they wish. For simplicity,
it will be assumed that the operations are carried out in the
order of increasing index, i.e., first $(m+1)$th party applies an
operation, then $(m+2)$th, etc. Let $s^{(k)}_\lambda$ and
$P^{(k)}_\lambda$ denote the scale factors and probabilities of
$k$th party's operation. Each local operation is a two outcome
general measurement where $\lambda=0$ corresponds to the failure
result with $s^{(k)}_0=0$ and $\lambda=1$ corresponds to the
success result. In order to satisfy the condition (i) of Theorem
\ref{thm:locop}, we should have $s^{(k)}_1\geq1$ and
$P^{(k)}_1=1/s^{(k)}_1$. Let $\mathbf{x}^{(k)}$ denote the state
after the successful operation of the $k$th party
($k=m+1,\cdots,p$). It is given as
\begin{equation*}
\mathbf{x}^{(k)}   = ( r_1y_1,\ldots, r_1 y_k,x_{k+1},\ldots,x_p )s^{(m+1)}_1\cdots s^{(k)}_1~.
\end{equation*}
As the components of all such vectors add up to $1$, the scale
factors are
\begin{equation}
  s^{(k)}_1 = \frac{r_1(y_1+\cdots+y_{k-1})+x_k+x_{k+1}+\cdots+x_p}{r_1(y_1+\cdots+y_k)+x_{k+1}+\cdots+x_p}~.
\end{equation}
Since $r_k>r_1$, we have $s^{(k)}_1>1$ and therefore the $k$th
operation can be carried out. The final state is obviously
$\mathbf{x}^{(p)}=\mathbf{y}$, and the probability of distillation
is given by
\begin{equation}
  P=P^{(m+1)}_1\cdots P^{(p)}_1=\frac{1}{s^{(m+1)}_1\cdots s^{(p)}_1}=r_1\quad,
\end{equation}
i.e., the upper bound
$\mathcal{P}_{\mathrm{bnd}}(\mathbf{x}\rightarrow\mathbf{y})$ has
been attained.

Finally, if $\mathbf{y}$ corresponds to a bipartite state between
parties $r$ and $s$ with concurrence $\mathcal{C}$, it can be
shown that $P\leq
\mathcal{P}_{\mathrm{bnd}}(\mathbf{x}\rightarrow\mathbf{y})=2\sqrt{x_rx_s}/\mathcal{C}$.
Considering the special case where $\mathbf{x}$ is also bipartite
entangled, it can be seen that this bound is also not tight.

\section{Conclusion \label{sec:conc} }

The essential mathematical tools for the systematic investigation
of all possible transformations of $W$-type states have been
obtained. These tools include the simplex $\mathcal{S}$, which is
extremely useful in the identification of LU equivalence classes
of the states. It appears that the LU equivalence classes of truly
multipartite states are represented by a single point in
$\mathcal{S}$, which considerably simplifies the analysis of the
transformations of such states. A complete characterization of
transformations that can be carried out by a single party is also
given. Finally, these tools are used for identifying all necessary
and sufficient conditions for deterministic transformations. An
upper bound for maximum distillation probability of arbitrary
multipartite states is also given. A complete characterization of
the probabilistic transformations of $W$-type states is still an
open problem. It is hoped that this article lays a good background
from which such problems and related questions about the
multipartite entanglement can be studied.

\end{document}